\documentclass[twocolumn,tighten,trackchanges]{aastex63}
\usepackage{mathtools}
\usepackage{txfonts} 
\usepackage{hyperref}
\usepackage{color}
\newcommand{\ud}{\mathrm{d}}

\defcitealias{PaperI}{EHTC~I}
\defcitealias{PaperII}{EHTC~II}
\defcitealias{PaperIII}{EHTC~III}
\defcitealias{PaperIV}{EHTC~IV}
\defcitealias{PaperV}{EHTC~V}
\defcitealias{PaperVI}{EHTC~VI}
\defcitealias{PaperVII}{EHTC~VII}
\defcitealias{PaperVIII}{EHTC~VIII}
\defcitealias{Narayan_2021}{N21}
\defcitealias{Gelles_2021}{G21}
\defcitealias{Deglinnocenti_1985}{LD85}

\newcommand{\npix}{$N_{\rm pix}$}%
\newcommand{\m}{M87*}
\newcommand{\s}{Sgr\,A*}
\begin{document}                                              

\turnoffeditone

\title{Supermassive Black Hole Spin Constraints from Polarimetry in an Equatorial Disk Model}
\shorttitle{Supermassive Black Hole Spin Polarimetry}

\author[0000-0002-7179-3816]{Daniel~C.~M.~Palumbo}
\affil{Center for Astrophysics $\vert$ Harvard \& Smithsonian, 60 Garden Street, Cambridge, MA 02138, USA}
\affiliation{Black Hole Initiative at Harvard University, 20 Garden Street, Cambridge, MA 02138, USA}

\begin{abstract}

The Event Horizon Telescope has released polarized images of the supermassive black holes Messier 87* (M87*) and Sagittarius A* (Sgr A*) accretion disks. As more images are produced, our understanding of the average polarized emission from near the event horizon improves. In this letter, we use a semi-analytic model for optically thin, equatorial emission near a Kerr black hole to study how spin constraints follow from measurements of the average polarization spiral pitch angle. We focus on the case of M87* and explore how the direct, weakly lensed image spiral is coupled to the strongly lensed indirect image spiral, and how a precise measurement of both provides a powerful spin tracer. We find a generic result that spin twists the direct and indirect image polarization in opposite directions. Using a grid search over model parameters, we find a strong dependence of the resulting spin constraint on plasma properties near the horizon. Grid constraints suggest that, under reasonable assumptions for the accretion disk, a measurement of the direct and indirect image spiral pitch angles to $\pm 5^\circ$ yields a dimensionless spin amplitude  measurement with uncertainty $\sigma_{|a_*|}\sim0.25$ for radially infalling models, but otherwise provides only weak constraints; an error of $1^\circ$ can reach $\sigma_{|a_*|}\sim0.15$. We also find that a well-constrained rotation measure greatly improves spin measurements. Assuming that equatorial velocity and magnetic field are oppositely oriented, we find that the observed M87* polarization pattern favors models with strong radial velocity components, which are close to optimal for future spin measurements.

\end{abstract}

\section{Introduction}

The mathematics of strongly lensed light around a black hole event horizon present a tantalizing universality; simple structures that neatly encode spacetime properties such as the mass, spin, and viewing inclination of the system \citep[see, e.g.][]{Johannsen_2010, Gralla_2020_null, Johnson_2020}. Of particular interest are predictions that may be accessible through near-term very-long-baseline interferometry (VLBI) experiments targeting the photon ring, such as the next-generation Event Horizon Telescope \citep[ngEHT,][]{Doeleman_2023} and Black Hole Explorer mission \citep[BHEX,][]{Johnson_2024}. The photon ring is the sharp image feature corresponding to nearly-orbiting light escaping to a distant observer, and is composed of sub-images indexed by the number $n$ of half-orbital windings undergone by photons before arriving at the observer. Universal structures emerge for large $n$.

The reality is that no structure observed in the next decade will have large $n$, or be otherwise universal -- that is, independent of astrophysical assumptions beyond zero optical depth. Simulations of the Messier 87* (\m{}) and Sagittarius A* (\s{}) accretion disks uniformly predict strong contributions from the first lensed sub-image of half-orbiting photons, the $n=1$ ring \citep{PaperV, Sgra_PaperV}. Indeed, studies in the Fourier domain suggest that the long-baseline signal already measured by the Event Horizon Telescope (EHT) may already have comparable contributions from the $n=0$ and $n=1$ images in total intensity, linear polarization, and circular polarization \citep{Broderick_2020, Palumbo_2023_b2vis, Shavelle_2024, Tamar_2024}. However, the contribution from the $n=2$ image is largely undetectable on baselines accessible now or in the near future due to the exponential suppression of subsequent sub-images, even in the absence of optical depth. 

We are left to reckon fully with the rich phenomenology of the $n=1$ image. In this letter, we focus on the polarimetry of the $n=0$ and $n=1$ images as a spin probe for \m{}, inspired by the spin imprint in alternating $n$ image polarization described in the universal regime by \citet{Himwich_2020}. As has been studied in detail in ray-traced images of general relativistic magnetohydrodynamic (GRMHD) simulations, polarimetric reversals across $n=0$ and $n=1$ are a ubiquitous feature of optically thin accretion disks with weak Faraday effects \citep{Ricarte_2021, Mosci_2021, Palumbo_2022, Emami_2023}. For much higher $n$, the azimuthal structure of linearly polarized spirals is a direct spin tracer (though spin information is erased at zero inclination), but for $n=0$ and $n=1$ images, this structure is very sensitive to astrophysical details. 

In order to study astrophysical uncertainties in spin inferences, we utilize \texttt{KerrBAM}, a semi-analytic accretion model for optically thin synchrotron emission from an axisymmetric, equatorial disk \citep{Palumbo_KerrBAM}; this model is a descendant of other recent toy models used for both \m{} and \s{} \citep{Narayan_2021, Gelles_2021}. We assume that the polarimetric spiral pitch angle coefficient, $\beta_2$, can be measured with some quantified error for both the $n=0$ and $n=1$ emission \citep{PWP_2020}. We predict the measured phases $\angle \beta_{2,0}$ and $\angle \beta_{2,1}$ across a large grid of models, and use self-fits to examine marginalized spin amplitude measurements under varying joint constraints on other model parameters, such as the plasma velocity orientation and the characteristic emission radius.

In Section \ref{sec:model}, we review the polarimetric spiral pitch angle and the accretion disk model, and demonstrate trends in polarized image structure that show strong spin signatures amid a rich astrophysical phenomenology. In Section \ref{sec:spin}, we show how spin constraints evolve with respect to measurement uncertainties and astrophysical parameters. We conclude with a discussion in Section \ref{sec:conclusion}.

\begin{figure*}[t!]
    \centering
    \includegraphics[width=\textwidth]{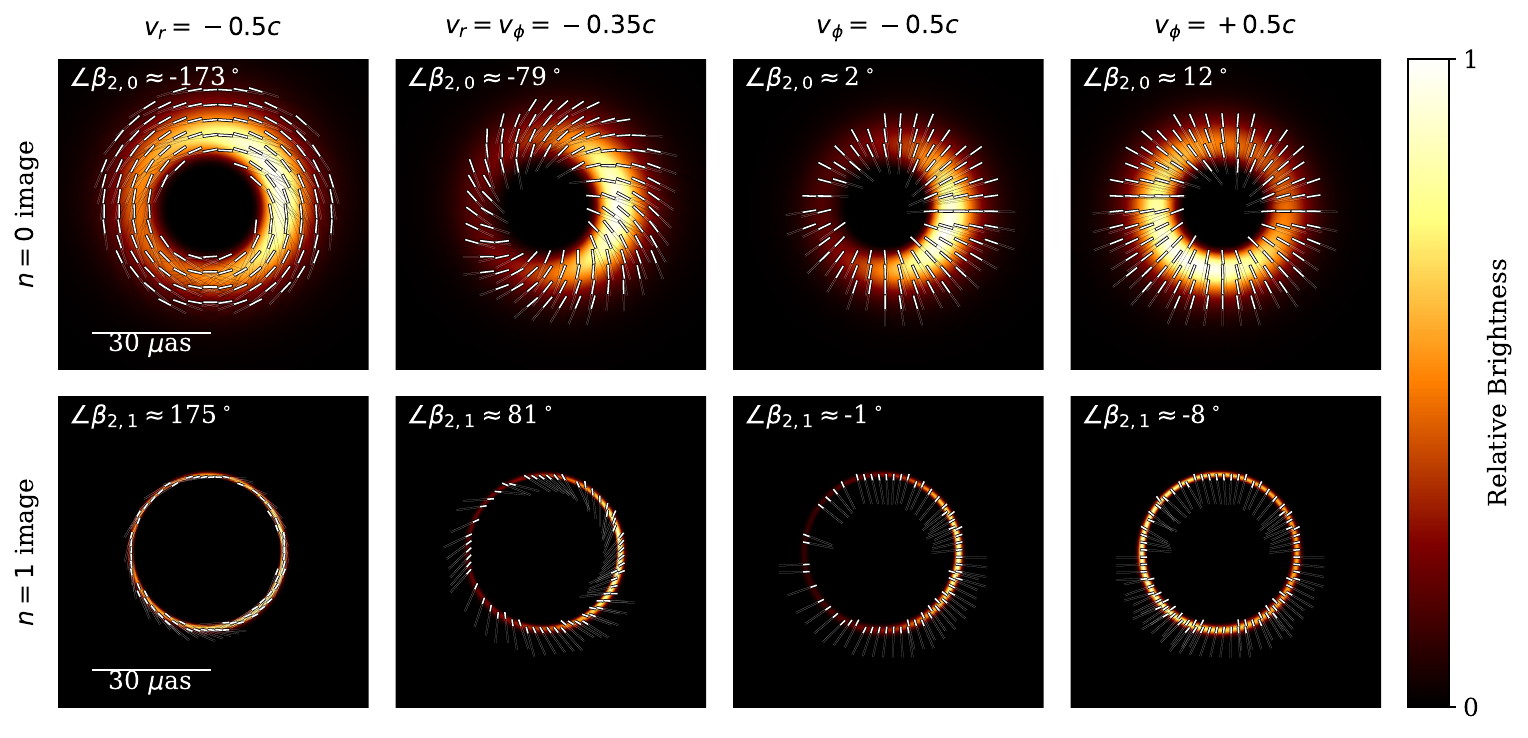}
    \caption{Example polarized images of the \texttt{KerrBAM} model with $\theta_o=17^\circ$, varying only the axisymmetric velocity orientation $\chi$. Velocities at top are the spatially uniform velocity in the ZAMO frame. The assumption that equatorial magnetic fields trail the plasma velocity vector causes $\chi$ to be the leading order term in determining the polarimetric spiral pitch angle $\angle \beta_2$.}
    \label{fig:model_intro}
\end{figure*}

\section{Polarimetric Model Trends}
\label{sec:model}

We begin with a brief review of the polarimetric spiral pitch angle, and then explore trends in the \texttt{KerrBAM} model space.

\subsection{Polarimetric spiral pitch angle}
We use the image-averaged rotationally symmetric polarization coefficient $\beta_2$, introduced in \citet{PWP_2020}. $\beta_2$ is one element of a larger radially-averaged azimuthal Fourier decomposition of the complex-valued linearly polarized image $P(\rho, \varphi)$, where $\rho$ is the image radius and $\varphi$ is the image azimuthal angle measured east of north:
\begin{align}
    \beta_m &=\dfrac{1}{I_{\rm tot}} \int\limits_{0}^{\infty} \int\limits_0^{2 \pi} P(\rho, \varphi) \, e^{- i m \varphi} \; \rho \mathop{\ud\varphi}  \mathop{\ud\rho},\\
    I_{\rm tot} &= \int\limits_{0}^{\infty} \int\limits_0^{2 \pi} I(\rho, \varphi) \; \rho \mathop{\ud\varphi} \mathop{\ud\rho}.
\end{align}
Here, $I(\rho,\varphi)$ is the corresponding total intensity image. Each coefficient $\beta_m$ is broadly proportional to the resolved fractional polarization.

The image-integration inside $\beta_2$ mixes the structures of the $n=0$ and $n=1$ images, causing drifts in the observed phase as a function of frequency as shifting optical and Faraday depths cause relative contributions to vary \citep{Palumbo_2024}. However, interferometers disentangle direct and indirect image structure; \citet{Palumbo_2023_b2vis} studied an example GRMHD simulation for \m{} and \s{} and found no frequency variation in the inferred values of $\beta_2$ taken from long-term averages of other observables, such as the corresponding Fourier quantity, $\breve{\beta}_2$, in baseline regimes clearly dominated by one sub-image or the other. Though we do not directly utilize $\breve{\beta}_2$, the construction is summarized in Appendix \ref{sec:app} for completeness.

In this work, it is sufficient to assume that repeated observations of \m{} on currently available EHT baselines permit an average measurement of $\beta_2$ for the direct image, while similar observations on longer (space-VLBI) baselines or at higher frequencies eventually yield a measurement of $\beta_2$ for the indirect image. Moreover, since the relative amplitudes of these two $\beta_2$ values will depend sensitively on astrophysics not modeled in this work such as different optical depths in the $n=0$ and $n=1$ image, we will discard their amplitude information, working only with the two phase quantities $\angle\beta_{2,0}$ for the direct image and $\angle\beta_{2,1}$ for the indirect image.

\subsection{The Semi-analytic Emission Model}

\begin{figure*}[t!]
    \centering
    \includegraphics[width=\textwidth]{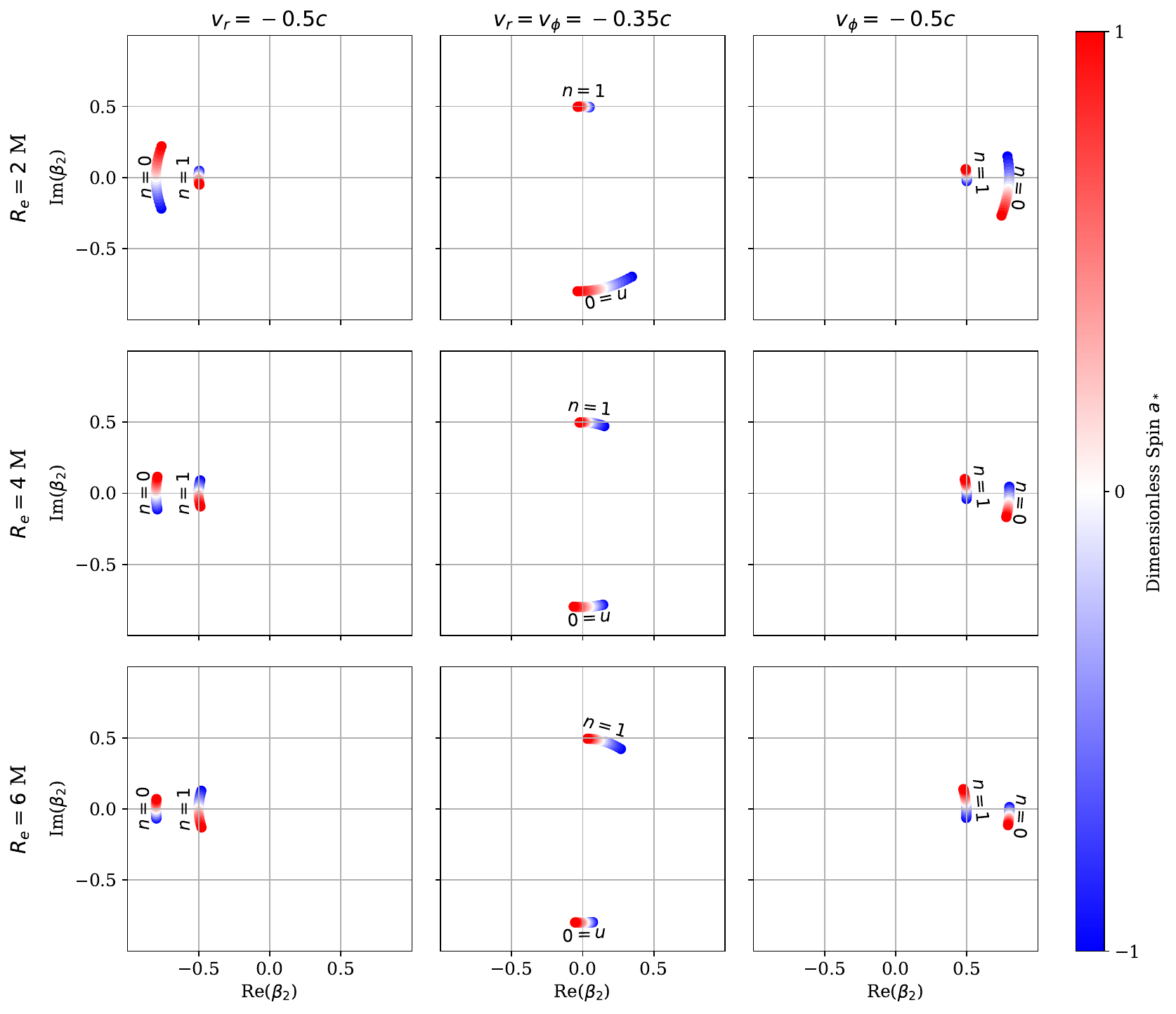}
    \caption{Evolution of the direct and indirect image $\angle \beta_2$ as a function of spin for various combinations of characteristic emission radius $R_e$ and axisymmetric velocity for a model wiith $\theta_o=17^\circ$. Negative values of $a_*$ (blue) are clockwise black hole spins on the sky; the black hole spin evidently rotates the direct image phase, $\angle\beta_{2,0}$ oppositely in the complex plane to the on-sky direction, while the indirect image phase $\angle\beta_{2,1}$ rotates in the same direction as the spin.}
    \label{fig:spin_drift}
\end{figure*}
The specification of the accretion disk in \texttt{KerrBAM} is optimized for speed and simplicity, aimed at producing reasonably accurate images of accretion systems dominated a small region in the equatorial plane. The model is described in full in \citet{Palumbo_KerrBAM}, but we summarize key properties here.

The ray tracing in the model is exact and semi-analytic using elliptic integrals to connect the observer screen to the emitting plane \citep{Rauch_1994, Dexter_2009, Gralla_2020_null}, and is carried out adaptively with increasing resolution for each sub-image $n$ given a specification of the FOV and pixel number $N_{\rm pix}$. The plasma specification is where most of the simplification happens; using the settings employed here, the model assumes a large number of plasma parameters are both axisymmetric and radially uniform:
\begin{enumerate}
    \item The plasma speed in the zero-angular-momentum-observer (ZAMO) frame, $\beta$.
    \item The equatorial pitch angle, $\chi$ of the axisymmetric velocity relative to the ZAMO radial unit vector in Boyer-Lindquist coordinates.
    \item The vertical polar angle, $\iota$, of the magnetic field vector, measured from the equatorial Boyer-Lindquist polar unit vector $-\hat{\theta}$ in the lab frame. The equatorial component of the magnetic field is assumed to be oriented oppositely to the fluid velocity, which is itself restricted to the midplane.
    \item The spectral index $\alpha_\nu$ of the fluid frame synchrotron spectrum.
\end{enumerate}
It bears emphasizing that assuming a radially uniform fluid velocity in the ZAMO frame and fluid frame magnetic field are both poor assumptions for GRMHD flows in full generality, especially if images indicate emission from distinct regions undergoing different dynamics. In addition, the assumption that the equatorial magnetic field and velocities should be opposed will not strictly hold for any magnetically dissipative flow, though the equatorial magnetic field seems to approximately obey this assumption in GRMHD \citep{Ricarte_2022}. Should the velocities and magnetic fields become decoupled, the primary determinant of image polarization structure will be the magnetic field, not the velocities; pinning them to one another reduces parameter volume while still allowing a separate polar magnetic field component. 

Absolute properties of the accretion disk such as the number density and temperature of electrons and the strength of the magnetic field are abstracted away; all radial structure not borne from the spacetime itself is absorbed into a radial profile $\mathcal{J}(r)$. We utilize a ring-shaped emissivity profile with a Gaussian cross-section:
\begin{align}
    \label{eqn:profile}
    \mathcal{J}(r; R_e, w) &= \exp\left[-4\log2 \left(\frac{r-R_e}{w}\right)^2\right].
\end{align}
Here, $R_e$ is the characteristic radius, and $w$ is the full-width at half-maximum of the Gaussian cross section. No emission is produced within the horizon.

%We utilize a double-sided power law specified by a characteristic emission radius $R_e$, and inner and outer indices $p_1$ and $p_2$:
% \begin{align}
%     \label{eqn:profile}
%     \mathcal{J}(r; R, p_1, p_2) &= \frac{(r/R_e)^{p_1}}{1+(r/R_e)^{p_1+p_2}}.
% \end{align}
This profile is scale-free; the total flux in resulting images is a free parameter. Due to our focus solely on the phases $\angle\beta_{2,0}$ and $\angle\beta_{2,1}$, the total flux of the model, and the relative fluxes in the $n=0$ and $n=1$ images, are irrelevant. However, the radial profile is still quite important, as it weights emission differently from distinct radii which produce different polarization structures.

Given the full specification of the plasma, the model predicts images given a mass (or mass-to-distance ratio), dimensionless spin, viewing inclination, and position angle of the spin axis. Once again, we may discard some parameters; $\beta_2$ is invariant under image rescalings and rotations, so only the spin $a_*$ and inclination $\theta_o$ matter. 

Given a fixed inclination $\theta_o < 90^\circ$, $a_*$ is a signed quantity where a positive sign indicates spin pointed toward the viewer (spinning counter-clockwise on the sky), while a negative spin indicates spin pointed away (spinning clockwise on the sky). In this convention, given evidence so far for the near-horizon clockwise rotation of \m{}, one would expect a $\chi<0$, $\theta_o<90^\circ$ and $a_*<0$. This convention is distinct from that often used in GRMHD studies by the EHT Collaboration, in which the sign of $a_*$ indicates retrogradeness relative to a large-scale accretion disk, and maintaining spin pointing away is obtained by taking $\theta_o \rightarrow 180-\theta_o$.

\begin{figure*}[ht]
    \centering
    \includegraphics[width=\textwidth]{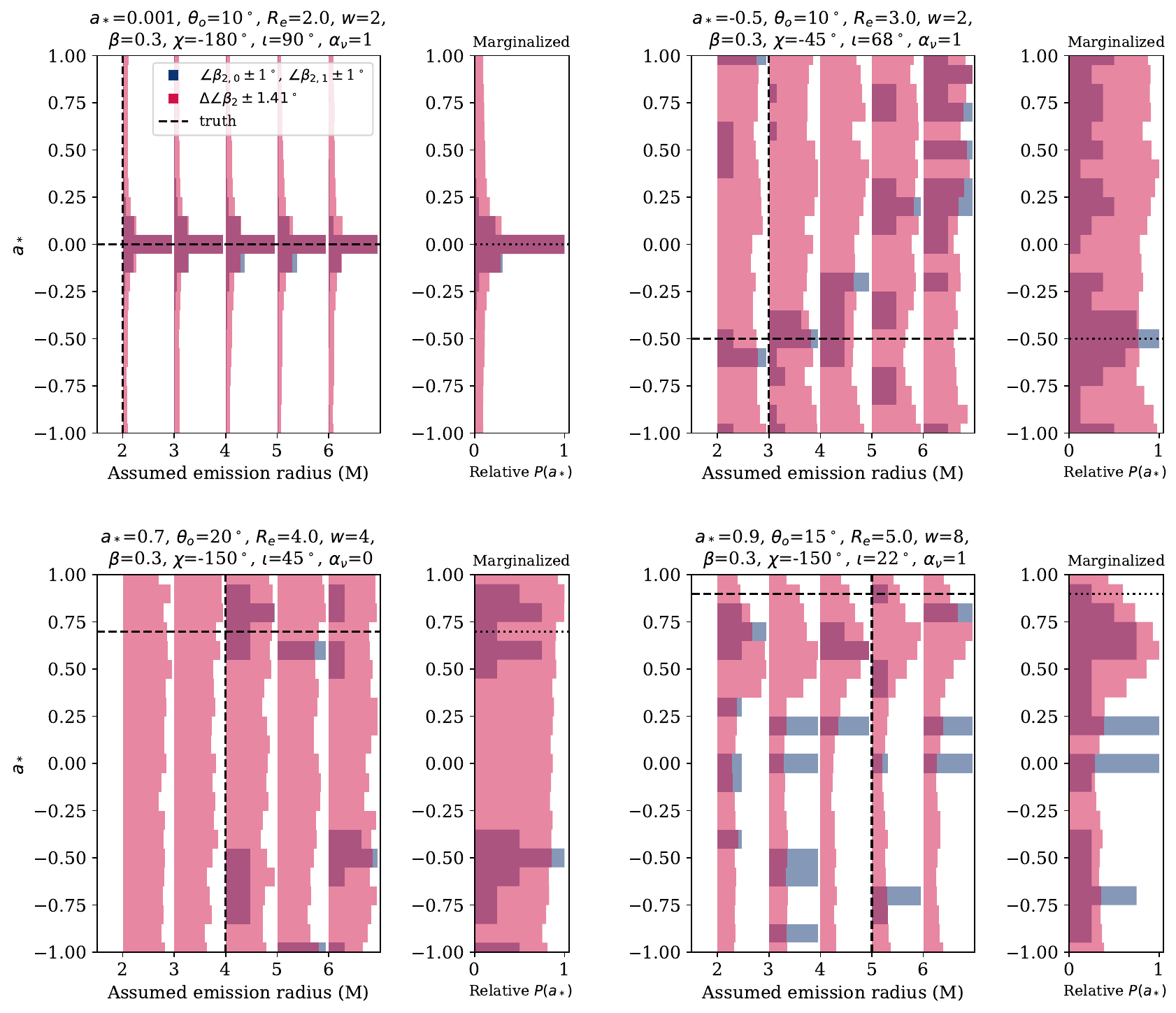}
    \caption{Example self-fit comparisons across the model grid in Table \ref{tab:params} for four true models with and without a known overall rotation measure, in the low-noise limit where the direct and indirect image $\angle \beta_2$ are each known to within a degree. For the top left radially infalling model, polarimetry is highly constraining even without a rotation measure constraint, whereas the lack of RM information destroys the spin constraint in all other cases. Each histogram shows the distribution of passing spins, broken down among emission radii at left and marginalized at right. Each histogram is normalized to produce unit peak probability density for visual clarity.}
    \label{fig:selffits}
\end{figure*}

Polarization is predicted pixel-wise by evaluating the synchrotron emissivity at geodesic impacts on the equatorial plane, and parallel transporting the polarization vector forward along the geodesic. Figure \ref{fig:model_intro} shows polarization patterns and values of the phase $\angle\beta_2$ for a model in which all model parameters are constant except for $\chi$, which is the leading order contribution to polarized morphology due to the assumption of magnetic fields opposing plasma velocities. Each column in the figure thus corresponds to a global rotation of the plasma velocity while holding speed constant. White tick marks indicate electric vector position angle (EVPA), the orientation of electric field oscillations on the sky. The ability of \texttt{KerrBAM} to naturally decompose sub-images is crucial to this study, and the near-complex-conjugation relation between direct and indirect image $\beta_2$ values found in  \citet{Palumbo_2022} is apparent.

Of interest is how these images vary when all plasma parameters are held constant but only spin is varied. Figure \ref{fig:spin_drift} shows the complex $\beta_2$ plane for nine models, where $\beta_{2,0}$ and $\beta_{2,1}$ are shown varying independently as a function of spin while holding all else constant. Radial distances in the figure between $n=0$ and $n=1$ $\beta_2$ values are only chosen for visual clarity. We find that the effect of spin is to twist the $n=0$ and $n=1$ values of $\angle\beta_2$ oppositely; $n=0$ rotates against the spin while $n=1$ rotates with it. Intuitively, this follows from the presence of terms in the Penrose-Walker constant that scale as $a_* p^z$ at the midplane (where $p$ is the photon momentum and $z$ is the vertical axis), which will cause a linear deformation in polarization in opposite directions for the $n=0$ and $n=1$ emission \citep{Walker_Penrose_1970}. Further, we find that the relative angular shift in each image depends sensitively on the astrophysical details; this figure varies emission radius $R_e$ along rows and axisymmetric velocity orientation $\chi$ along columns. 

We also see that whether the $n=0$ or $n=1$ polarization spiral pitch angle varies more with respect to spin depends on the emission radius. When the emission radius $R_e$ is small, the direct image polarization varies widely over values of $a_*$, whereas when the emission radius is large, the $n=1$ image carries most of the variation. This effect corresponds approximately to whether $\angle\beta_{2,1}-\angle\beta_{2,0}$ is positive or negative modulo $2\pi$, and the tipping point in $R_e$ appears to be close to the photon sphere. These results capture a rotationally symmetric manifestation of a similar effect in Figure 3 of \citet{Connors_et_al_1980}.

We see by inspection that measuring a particular pair of $\angle\beta_{2,0}$ and $\angle\beta_{2,1}$ does not yield a single spin value without characterizing the rest of the properties of the plasma. This figure is just slice through a vast parameter volume; further, \texttt{KerrBAM} treats only one possible polar angle among many possible contributing surfaces in the emission (as are treated in a more complete geometric model in \citet{Chang_2024}).

We now move on to a broad survey of the model parameter space to see how individual polarimetric measurements translate into spin constraints.

\section{Spin Inference and Uncertainty}
\label{sec:spin}
Though \texttt{KerrBAM} contains a Bayesian inference framework for posterior estimation of model parameters in VLBI data applications, this approach is prohibitively expensive for exploring the evolution of data constraints over a large set of true parameters. Rather, by isolating the image-domain feature $\angle\beta_2$ from each of the $n=0$ and $n=1$ images, we may more straightforwardly estimate rough constraints in parameter space by dictating a proposed measurement and uncertainty for each value, and applying these as a hard constraint in a grid of \texttt{KerrBAM} models.

We begin by evaluating a large product space of  $n=0$ and $n=1$ sub-images of the semi-analytic model. The model values are given in Table \ref{tab:params}. Each parameter range is chosen to qualitatively explore many different determinants of polarimetric morphology, particularly in the context of the \m{} problem in which the inclination of the spin axis is well-constrained. The plasma velocities range from sub-Keplerian ($\beta=0.3$ at $R=2$M) to super-Keplerian ($\beta=0.9$ at $R=6$M). We compute three pieces of information across this grid: $\angle\beta_{2,0}$, $\angle\beta_{2,1}$, and $\Delta\angle\beta_2$, defined as the signed rotation between the two coefficients:
\begin{align}
    \Delta\angle\beta_2 &\equiv {\rm arg}(\beta_{2,1}\beta_{2,0}^*).
    \label{eq:delta}
\end{align}
$\Delta \angle\beta_2$ discards the absolute phase of the EVPA, and is therefore all that is available for interpretation an unknown Faraday screen coherently rotating the EVPA of the $n=0$ and $n=1$ images by the same amount. Note that this discounts internal Faraday rotation that would manifest as a non-gravitational differential rotation between the $n=0$ and $n=1$ emission, which are self-consistently predicted by GRMHD. \citet{Palumbo_2023_b2vis} found no frequency dependence in the interferometric polarization structure individually dominated by the $n=0$ or $n=1$ ring, suggesting that these differential effects are small. Nonetheless, these concerns are particularly relevant for the Galactic Center, which has a large and rapidly varying rotation measure (RM) even on EHT scales at 230 GHz \citep{SgrA_Paper_VII, SgrA_Paper_VIII, Wielgus_2024}.

\begin{table}[t]
\centering
\caption{
%Parameter symbols and prior ranges
Grid search model parameters. In total, \edit1{1,008,000} models (parameter combinations) are considered.
}
\label{tab:params}
\begin{tabular}{ccc}
\hline
\hline
\textbf{Ray Tracing Parameters} & \textbf{Symbol} & \textbf{Values} \\
\hline
Dimensionless Spin & $a_*$ & -1, -0.9, ..., 0.9, 1\\
\hline
Observer Inclination $({}^\circ)$ & $\theta_{\rm o}$ & \edit1{10, 15, 20, 25, 30}\\
\hline
Field of View ($\mu$as along edge) & FOV & 96 (Fixed)\\
\hline
Number of Pixels (along edge)  & \npix & 192 (Fixed)\\
\hline
\hline
\textbf{Fluid Parameters} & & \\
\hline
Fluid Speed (fraction of $c$) & $\beta$ & 0.3, 0.5, 0.7, 0.9\\
\hline
Equatorial Fluid Velocity Angle $({}^\circ)$ & $\chi$ &  \edit1{-180, -165, ..., 165} \\ %-180, -157.5, ..., 157.5 \\
\hline
Vertical Magnetic Field Angle $({}^\circ)$ & $\iota$ & 0, 22.5, 45, 67.5, 90 \\
\hline
Spectral Index & $\alpha_\nu$ & 0, 1\\
\hline
\hline
\textbf{Emission Profile Parameters} & &\\
\hline
Characteristic Radius (M) & $R$ & 2, 3, 4, 5, 6\\
\hline
Full-width at half maximum (M) & $w$  & 2, 4, 8 \\
\end{tabular}
\end{table}

\subsection{Constraints from Self-Comparisons}
\label{subsec:self-fit}

Figure \ref{fig:selffits} shows the spin constraints resulting from comparing the full model grid to single models, allowing a $1^\circ$ uncertainty on the direct and indirect spiral pitch angles $\angle\beta_{2,0}$ and $\angle\beta_{2,1}$. Four example models are shown, varying most of the model parameters. For each model, we compute the $n=0$ and $n=1$ spiral pitch angle, and construct a ``passing'' subset of models based on comparison from these pitch angles to the model grid pitch angles, subject to the specified uncertainty as a hard cut. We examine two cases: one in which the absolute phases are known (corresponding to a known RM), and another in which the RM is unknown, in which case only $\Delta\angle\beta_2$ is utilized.

We observe two striking features. First, that the spiral pitch angle measurements can be either strongly constraining (as in the left two models) or almost completely unconstraining (as in the right two models) \edit1{of the spin amplitude} depending on the underlying true models. That is, some positions in plasma parameter space are more amenable to spacetime measurements than others. Second, we see that the absence of a rotation measure constraint seems to destroy all spin information in the polarimetric signal, with the exception of the radially infalling model (top left), though the uncertainty around $a_*=0$ does widen in the absence of the RM constraint. 

Notably, the weak spin constraints in the right two columns are present despite the very low error on the polarimetric spirals; realistic errors will likely be larger.

\subsection{Spin Uncertainty Evolution over Parameter Space}

\begin{figure}[t!]
    \centering
    \includegraphics[width=0.47\textwidth]{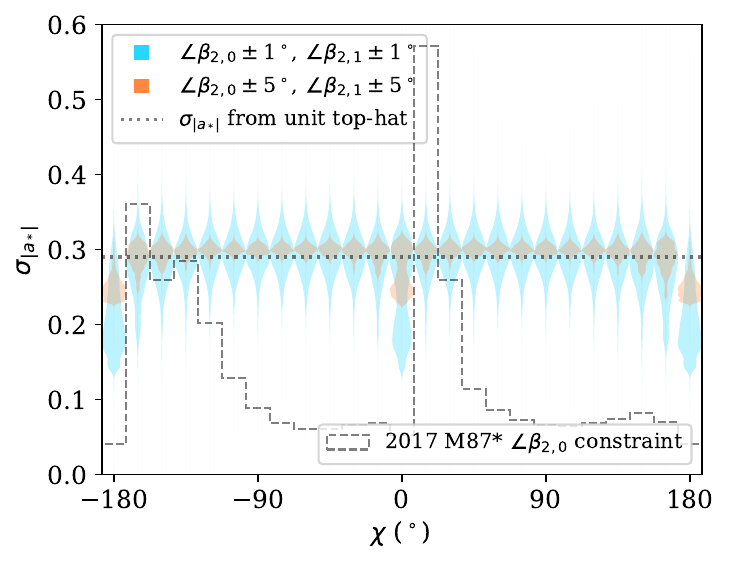}
    \caption{Survey of spin amplitude uncertainties resulting from filtering the full grid search according to either a $1^\circ$ or $5^\circ$ error on each of $\beta_{2,0}$ and $\beta_{2,1}$. Violins show the distribution of spin errors over the space of underlying true models, decomposed along the horizontal axis according to the axisymmetric velocity pitch angle $\chi$ of the true model. Models with radially infalling ($\chi=180^\circ$) or outflowing $(\chi = 0^\circ$) material show the strongest sensitivity to spin at fixed values of the measurement uncertainty. }
    \label{fig:uncertainty}
\end{figure}
Just as each model in the parameter space predicts a single value for each of $\angle\beta_{2,0}$ and $\angle\beta_{2,1}$, each model also predicts a spin constraint given the uncertainty on each spiral pitch angle. We may characterize the variation of spin-constraining power over the model grid by conducting fixed-uncertainty searches, in which permitted spins for each measured true model are used to estimate the resulting spin amplitude uncertainty, $\sigma_{|a_*|}$. That is, we first choose uncertainties $\sigma_{\angle\beta_{2,0}}$ and $\sigma_{\angle \beta_{2,1}}$; then, for every model in the grid, we compute $\angle\beta_{2,0}$ and $\angle\beta_{2,1}$, then determine which models in the grid predict sub-image polarimetric phases within each chosen $\sigma$ of the originally proposed model. This collection of ``passing'' models provides an estimate of the spin amplitude uncertainty for each underlying true model.

Of particular interest is the evolution in spin uncertainty over $\chi$ under our assumption that the equatorial magnetic field follows the plasma velocity. As was apparent in Figure \ref{fig:spin_drift}, changing $\chi$ differentially rotates the spiral pitch angle in the $n=0$ and $n=1$ image. Values of $\chi=0^\circ$ and $\chi=-180^\circ$ minimize the difference between the two sub-image values with $a_*=0$. Evidently from Figure \ref{fig:selffits}, these regimes present strong spin sensitivity from polarization structure.

The polarimetric observations of \m{} in \citet{PaperVII} were found to be consistent with a range of values of image-integrated $\angle\beta_2$, between $-163^\circ$ and $-127^\circ$. We may extract loose preferences on $\chi$ in the equatorial model by assuming that EHT observations probe only $n=0$ emission, in which case we compare the model grid to the constraint $-163^\circ < \angle\beta_{2,0} < -127^\circ$.

Figure \ref{fig:uncertainty} shows the result of the uncertainty survey with two values of the assumed measurement uncertainty, overplotted with the \m{} preferred values of $\chi$. Notably, \m{} does not eliminate any values of $\chi$ from polarimetric spiral pitch angle alone, though it clearly prefers values in which $-180^\circ<\chi<-90^\circ$, corresponding to clockwise motion with radial infalling velocity components, consistent with the interpretation in \citet{PaperVIII}. The violins represent the distribution of spin amplitude measurement error over the space of true models; for most values of $\chi$, these violins are centered at $\sigma_{|a_*|} \approx 0.3$. This value may seem promising at first, but we note with the overlayed dotted line that this corresponds closely to the standard error of a unit top hat, that is, a spin amplitude distribution that is uniform on the interval $[0,1]$. Portions of the model space yielding worse values of $\sigma_{|a_*|}$ than 0.3 correspond to spin constraints that happen to be multimodal with larger variance than a uniform distribution.

We observe that the values of $\chi$ for which the uncertainty $\sigma_{|a_*|}$ is minimized at each value of the measurement error are those with predominantly radial motion, that is, $\chi = -180^\circ$ and $\chi=0^\circ$. Though the \m{} constraint disfavors these particular values, the constraint prefers values quite near these peaks.

% \subsection{Observational Requirements from GRMHD Variation}
% We have learned from the self-constraints of the semi-analytic model that particular levels of certainty around image parameters such as the centroid and $\angle\beta_2$ translate into spin constraints. In applications to time-evolving accretion disks, these are each time-varying parameters for which \texttt{KerrBAM} only approximates the mean.

% In order to determine the number of times future instruments must measure these parameters in order to measure spin, we may demand that the error on the mean of any given parameter be equal to a requisite level of uncertainty in the semi-analytic model, $\sigma_{\rm SA}$:
% \begin{align}
%     \frac{\sigma_{\rm GRMHD}}{\sqrt{n}}&=\sigma_{\rm SA},\\
%     n &= \left(\frac{\sigma_{\rm GRMHD}}{\sigma_{\rm SA}}\right)^2.
% \end{align}
% We will take $\sigma_{\rm SA}$ from the analysis in \ref{subsec:self-fit}, and measure $\sigma_{\rm GRMHD}$ from model images.

% \begin{align}
%     \sigma_{{\rm arg}(\tilde{I})} &= u \sigma_x,\\
%     &\approx 7^\circ \times \left(\frac{u}{25 {\rm G}\lambda} \right)\left(\frac{\sigma_x}{1 \mu{\rm as}}\right)
% \end{align}

\section{Conclusion}
\label{sec:conclusion}

We have extended a simple equatorial model for optically thin emission in the Kerr equatorial plane to predict polarized emission. We have used this model to build a large grid of images over a parameter space representing both reasonable and unreasonable structures for the near-horizon accretion disk in \m{}. We have examined a particular image-integrated quantity, the polarization spiral pitch angle $\angle\beta_2$, for the direct ($n=0$) and indirect ($n=1$) image on the observer screen, and found that spin has opposite effects on the direct and indirect image polarization, twisting polarization through parallel transport with opposite handedness in alternating sub-images. We examined model space constraints resulting from uncertain measurements of each sub-image value of the spiral pitch angle, and found that the exact position in plasma parameter space can have a strong impact on the constraining power of the polarimetric measurement. We found that, under the assumption of antialigned equatorial velocities and magnetic fields, observations of \m{} favor velocities directed more radially than azimuthally (though they disfavor purely radial motion), which we find to be some of the most promising models for spin constraints from polarimetry.

These results are notable in part because they show how little power polarization information has in measuring spin in the absence of a prior on the plasma. For example, in GRMHD \citep{PWP_2020} and in semi-analytic magnetospheric prescriptions \citep{Chael_2023, Hou_Huang_2024}, \edit1{polarimetric morphology, in particular} $\angle\beta_2$ was found to provide a strong, nearly linear relationship with respect to spin. In this work, though our model is much simpler, we permit a wider range of fluid velocities and magnetic field geometries, allowing us to paint plasmas into the Kerr spacetime that erase much of the relationship between polarization and spin. Our results can thus be treated as an examination of spin constraints emerging purely from parallel transport with minimal imprint from black hole-accretion disk coupling in the magnetic field.

\edit1{Nonetheless, one may still use GRMHD simulations to estimate observation requirements in light of the uncertainty survey in this work. In order for average polarization structure to be used to measure spin, the phases $\angle\beta_{2,0}$ and $\angle\beta_{2,1}$ themselves must have a small error on the mean. We may require that the error on the mean polarization in an observation be smaller than a requisite error, $\sigma_{\rm req}$. GRMHD simulations allow us to estimate the intrinsic deviation, $\sigma_{\rm int}$, for which we must observe a number of times $N$ such that
\begin{align}
    \frac{\sigma_{\rm int}}{\sqrt{N}}&<\sigma_{\rm req},\\
    \rightarrow N_{\rm min} &\equiv \left(\frac{\sigma_{\rm int}}{\sigma_{\rm req}}\right)^2.
\end{align}
\citet{Palumbo_2022} examined distributions of $\beta_{2,0}$ and $\beta_{2,1}$ for a library of GRMHD simulations, and found a wide variety of distributions. However, for the phase in a typical magnetically arrested disk, $\sigma_{\rm int} \approx 30^\circ$, with comparable variation in $n=0$ and $n=1$ phases. Taking $\sigma_{\rm req} = 5^\circ$ to match  the less constraining values in Figure \ref{fig:uncertainty}, we find the requisite number of observations $N_{\rm min} \approx 36$. As repeated observations of \m{} and \s{} occur, empirical determinations of $\sigma_{\rm int}$ will become available. However, given the current EHT observation cadence, observing this many independent accretion realizations would take decades, and may not reach $n=1$ spatial scales in \m{} even at 345 GHz. Thus, the high observing cadence of the ngEHT and BHEX and the long baselines uniquely afforded by BHEX are necessary to guarantee sensitivity to time-averaged photon ring features.}

EHT theoretical constraints so far suggest that retrogradely spinning magnetically arrested disks excel at reproducing the image properties of \m{} \citep{PaperV, PaperVIII, PaperIX}. These models are notable in our context  because they produce nearly radial velocity distributions, as material loses angular momentum due to frame dragging as it approaches the horizon. As shown in the second column of Figure 2 in \citet{Palumbo_2022}, these models also produce values of $\angle\beta_{2,0}$ and $\angle\beta_{2,1}$ that are close to each other, similar to what we obtain in Figure \ref{fig:spin_drift} for the radial inflow models which are optimal for spin inference. Bondi accretion \citep{Bondi_1952} and other radial infall accretion scenarios \citep[see, e.g.][]{Falcke_2000, Narayan_2019_shadow} are also ideal for the spin measurement we describe. \edit1{However, it bears emphasizing that the detailed error budgets found in this paper will not extend to more general three-dimensional emitter morphologies, including in GRMHD. Though KerrBAM can reproduce Stokes $I$ morphology in magnetically arrested disk images from GRMHD \citep{Palumbo_KerrBAM}, a more thorough analysis of spin effects on polarization from off-equatorial emitters is necessary, extending work in \citet{Chang_2024}.}

Looking ahead to future observationally-minded studies of the photon ring, we expect much of the rich structure inflicted on images by the spacetime to only be accessible in the non-universal lensing regime. The polarized signatures analyzed here provide a unique example in which, in the face-on limit, all spin information in lost in the universal regime (see Equation 37 in \citet{Himwich_2020}). We demonstrate here the power of non-universal structure, with which observers and theorists alike will have no choice to reckon in ngEHT and BHEX observations, \edit1{provided they routinely detect long-baseline polarization, as they are currently envisioned to \citep{Doeleman_2023, Johnson_2024, Lupsasca_2024}}. Holistically, we encourage the application of semi-analytic models which model both the emitting geometry and the spacetime, to build a full and proper treatment of these non-universal features.

\acknowledgments{We thank Michael Johnson, Ramesh Narayan, and Dominic Chang for many helpful discussions in the preparation of this manuscript. We are grateful to our internal EHT referee for their thoughtful feedback on the manuscript. We also thank our journal referee for helpful comments on observational and systematic uncertainties in polarimetric morphology. We acknowledge financial support from the National Science Foundation (AST-2307887). This work was supported by the Black Hole Initiative, which is funded by grants from the John Templeton Foundation (Grant 62286) and the Gordon and Betty Moore Foundation (Grant GBMF-8273) - although the opinions expressed in this work are those of the author(s) and do not necessarily reflect the views of these Foundations. }

\appendix

\section{Measuring Polarimetric Pitch Angle with Interferometry}
\label{sec:app}
Though this manuscript does not operate on Fourier quantities, it is useful to imagine how the direct ($n=0$) and indirect ($n=1$) image polarization morphologies might be disentangled by an interferometer using Fourier quotients as in \citet{Palumbo_2023_b2vis}. The quantity $\breve{\beta}_2$ corresponds to a rotation and projection of interferometric quantities that contain the same information as the interferometric fractional polarization $\breve{m}$. Beginning with the interferometric Stokes $Q$ and $U$ quantities $\tilde{Q}(u,v)$ and $\tilde{U}(u,v)$, with $\theta = \arctan(u/v)$, we construct $E$ and $B$ mode polarization in the Fourier domain:
\begin{gather}
\label{eq:rot}
 \begin{bmatrix} \tilde{E}(u,v) \\ \tilde{B}(u,v) \end{bmatrix}
 =
  \begin{bmatrix}
   \cos 2 \theta &
   \sin 2 \theta \\
   -\sin 2 \theta &
   \cos 2 \theta 
   \end{bmatrix}
   \begin{bmatrix}
   \tilde{Q}(u,v) \\ \tilde{U}(u,v)
   \end{bmatrix},
\end{gather}
We then divide by the Stokes $I$ visibility to form a polarimetric closure, and project out the real part to target symmetric structures:
\begin{align}
    \breve{e}(u,v) &\equiv \frac{\tilde{E}(u,v)}{\tilde{I}(u,v)},\\
    \breve{b}(u,v) &\equiv \frac{\tilde{B}(u,v)}{\tilde{I}(u,v)},\\
    \breve{\beta}_2(u,v) &\equiv{\rm Re}(\breve{e}(u,v)) + i \, {\rm Re}(\breve{b}(u,v)).
    \label{eq:beta2}
\end{align}
Collections of measured polarimetric closures can be averaged coherently over long periods of time. Various track-averaging and annular-bin averaging strategies have been proposed \citep{Shavelle_2024, Tamar_2024}; for our purposes, it is sufficient that some suitable averaging scheme is taken such that, for a given sub-image-dominated regime in $(u,v)$:
\begin{align}
    \angle\beta_2 &\approx {\rm arg}\left(\left\langle \breve{\beta}_2\right\rangle\right),\\
    \sigma_{\angle\beta_2} &\approx \frac{\sigma_{|\left\langle \breve{\beta}_2\right\rangle|}}{|\left\langle \breve{\beta}_2\right\rangle|}.
\end{align}
These formulae correspond to the high signal-to-noise ratio limit for complex random variables; in a more general treatment of some set of Fourier data $\tilde{I}$, $\tilde{Q}$, and $\tilde{U}$, these formulae need not be assumed to still place meaningful constraints on polarized structure.

\bibliography{main}

\end{document}